\documentclass[twocolumn,showpacs,preprintnumbers,amsmath,amssymb,prl]{revtex4-1}
\usepackage{graphicx}
\usepackage{dcolumn}
\usepackage[tight]{subfigure}
\usepackage{amsmath}
\usepackage{verbatim}
\usepackage{color}
\usepackage{bm}
\newcommand{\fracs}[2]{{\textstyle \frac{#1}{#2}}}
\renewcommand{\vec}[1]{\mathbf{#1}}
\newcommand{\vecg}[1]{\pmb{ #1}}                 
\newcommand{\veco}[1]{ \hat{ \mathbf{#1}} }
\newcommand{\tens}[1]{\mathbf{#1}}
\newcommand{\tenso}[1]{\hat{\mathbf{#1}}}
\newcommand{\bra}[1]{\langle #1|}
\newcommand{\ket}[1]{|#1 \rangle}
\newcommand{\braket}[1]{\langle #1 \rangle}

\newcommand{\Tr}{\text{Tr}}
\newcommand{\Fig}[1]{Fig.~\ref{#1}}
\newcommand{\Eq}[1]{Eq.~(\ref{#1})}
\newcommand{\eq}[1]{(\ref{#1})}

\begin{document}
\title{Spin quadrupoletronics: moving spin anisotropy around}
\author{Michael Baumg\"artel$^{(1,2)}$}
\author{Michael Hell$^{(1,2)}$}
\author{Sourin Das$^{(3,1,2)}$}
\author{Maarten R. Wegewijs$^{(1,2,4)}$}
\affiliation{
  (1) Institut f\"ur Festk{\"o}rperforschung,
      Forschungszentrum J{\"u}lich, 52425 J{\"u}lich,  Germany \\
  (2) JARA- Fundamentals of Future Information Technology\\
  (3) Department of Physics and Astrophysics,\\
      University of Delhi, Delhi 110 007, India  \\
  (4) Institut f\"ur Theoretische Physik A,
      RWTH Aachen, 52056 Aachen,  Germany
}
\begin{abstract}
We show that spin anisotropy can be transferred to an isotropic system by \emph{transport} of spin quadrupole moment.
We derive the quadrupole moment current and continuity equation and
study a high-spin valve structure consisting of two ferromagnets coupled to a quantum dot probing an impurity spin.
The quadrupole back-action on their coupled spin results in spin torques and anisotropic spin relaxation which do not follow from standard spin current considerations.
We demonstrate the detection of the impurity spin by charge transport and its manipulation by electric fields.
\end{abstract}
\pacs{
  85.75.-d
  73.63.Kv,
  85.35.-p
}
\maketitle
The field of spintronics is driven by the desire to use the intrinsic dipole moment of the electron,
resulting from its spin, as an information carrier.
The investigation and design of spintronic devices has made great progress in understanding
the accumulation of spin dipole moments,
their manipulation by, e.g., current induced spin torques,
and their readout by electrical transport measurements.
Recently these studies have been extended to  molecular scale quantum dot (QD) devices~\cite{Sahoo05},
in particular the predicted~\cite{Koenig03} interplay of virtual tunneling, spin polarization and Coulomb interaction has been experimentally demonstrated in spin valve structures~\cite{Pasupathy04kondo,*Hauptmann08}.
Furthermore, the importance of \emph{intrinsic spin anisotropy}, 
induced by strong spin-orbit interaction, has been demonstrated in
 several measurements of transport through single magnetic molecules~\cite{Heersche06,*Jo06,*Osorio10}
and its various effects have been studied theoretically~\cite{Elste06,*Gonzalez07,*Misiorny09,Sothmann10}.
Using various experimental techniques, it was shown that the intrinsic magnetic  anisotropy can even be controlled by
atomic STM manipulation~\cite{Hirjibehedin06,*Hirjibehedin07,*Otte08,*Otte09},
mechanical straining~\cite{Parks10}
and reversible charging of the molecule controlled by a gate voltage~\cite{Zyazin10}.
The spin anisotropy is of central importance to molecular scale spin manipulation, as it can provide an energy barrier preventing unwanted spin reversal.
This has been a key motivation in the field of single-molecule magnetism and
proposals for quantum computing with magnetic molecules~\cite{Leuenberger01} also rely on spin anisotropy.

Spin anisotropy of a quantum state can be quantified by the average of the quadrupole moment tensor operator
\begin{align}
  \hat{Q}_{ij} =
    \fracs{1}{2} \left( \hat{S}_i \hat{S}_j
               + \hat{S}_j  \hat{S}_i \right)
  - \fracs{1}{3} \veco{S}^2 \delta_{ij}
  \label{eq:Q}
\end{align}
where $\hat{S}_i$ is the $i=x,y,z$ component of the spin operator $\veco{S}$.
This operator is non zero only for spin values $\ge 1$ and traceless.
Its components appear in the spin Hamiltonians describing the intrinsic spin anisotropy of magnetic molecules.
In this Letter we  show that a \emph{nonequilibrium spin-quadrupole moment} (SQM) can be induced in a QD, which by itself is spin \emph{isotropic}, by connecting it to ferromagnets.
We show that the SQM as a transport quantity obeys a continuity equation
 and locally affects the QD charge and spin and thereby the measurable charge current.
Spin anisotropy stored elsewhere can thus be \emph{transferred} to a place of interest by quantum electron transport.
Similar to spin splittings induced by tunneling~\cite{Pasupathy04kondo},
the effect of this anisotropy can be large compared to the intrinsic one
 when coupling strongly to the ferromagnets.
Moreover, the accumulated SQM can be controlled by electrical fields.
Accumulation of a local SQM requires a high-spin quantum dot with $S=1$ (or larger) in one of its charge states
and  a finite transport voltage.
We illustrate the importance of SQM currents for the charge and spin transport across an archetypal model of such a QD ``high-spin valve'' structure:
two ferromagnets coupled to a single orbital level side-coupled to an impurity spin 1/2.
In the nonequilibrium stationary state  this QD system has a finite SQM in the high-spin charge state,
in addition to noncollinear spins in the two successive charge states.
Throughout the paper we set $e=\hbar=k_\text{B}=1$
and distinguish between an operator $\hat{A}$ and its average $A=\braket{\hat{A}}$ only when needed.
\par
\emph{High-spin valve model.}
In QDs close to resonance, two adjacent discrete charge states dominate the transport.
The simplest model Hamiltonian of a QD with nonzero spin in both these charge states~\cite{Kiesslich09} is
\begin{align}
  H =  \varepsilon \hat{N} + U \hat{N}_\uparrow \hat{N}_\downarrow - J \, \veco{s} \cdot \veco{s}'
  .
\end{align}
Here $\varepsilon$ denotes the energy of an orbital with occupation operators $\hat{N}=\sum_\sigma \hat{N}_\sigma$, 
$\hat{N}_\sigma= d^{\dagger}_\sigma d_\sigma$ with $\sigma=\uparrow,\downarrow$
and spin
$\veco{s}=\fracs{1}{2} \sum_{\sigma \sigma'} d^{\dagger}_{\sigma} \vecg{\sigma}_{\sigma \sigma'} d_{\sigma'}$
with Pauli-matrix vector $\vecg{\sigma}$.
The electron on the orbital is side-coupled to a spin 1/2 $\veco{s}'$  with isotropic ferromagnetic exchange $J>0$.
This may represent, e.g., a QD coupled to a magnetic impurity~\cite{Heersche06b} or
a fullerene~\cite{Grose08,*Eliasen10},
 an asymmetrically gated double QD
~\cite{Kikoin02},
or even a hyperfine coupled single nuclear spin in a molecule~\cite{Ishikawa05Ho}.
Notably, both sign and magnitude of $J$ can be tuned in nanojunctions~\cite{Roch09,*Osorio10}.
We consider the limit $U \gg J \gg V_b, T \gg \Gamma_r$
where $V_b$ is the bias voltage and $\Gamma_r$ the tunnel rate.
Then the electron number $N$ is restricted by Coulomb blockade to 0 or 1
and the $N=1$ singlet state can be neglected due to the strong exchange.
Keeping only the ground states of the coupled orbital - impurity system with spin 1/2 ($N=0$) and 1 ($N=1$), respectively and energy difference $\epsilon=\varepsilon - J/4$, we obtain the simplest realization of an isotropic high-spin QD.
The orbital is tunnel coupled to noncollinearly polarized ferromagnets (FM)
,
$
  H_\text{F}  =  \sum_{r k\tau} \epsilon_{rk\tau} \, c_{r k \tau}^{\dagger}c_{r k \tau}
$
,
with a constant, spin polarized density of the states $k$ (DOS) $\sum_k \delta(\epsilon_{rk\tau} -\omega) \approx \nu_{r\tau}$
where $\tau=\uparrow,\downarrow$ refers to the spin of the electrons quantized along the polarization axis  ${\vec{n}}_r$ of the respective electrode $r=L,R$.
We let the length of the vector $\vec{n}_r$ denote the relative polarization of the density of states of the FM:
$|\vec{n}_r|=(\nu_{r\uparrow} - \nu_{r\downarrow})/(\nu_{r\uparrow} + \nu_{r\downarrow})$
where
$\nu_{r \uparrow} > \nu_{r \downarrow}$ by the definition of $\vec{n}_r$.
The tunnel coupling to the FMs is accounted for by
$
  H_\text{T}  =  \sum_{rk\sigma\tau} t_{r \sigma \tau}\, d_\sigma^\dagger c_{r k \tau} + \text{h.c.}
$
where $\sigma$ refers to the spin of electrons quantized along an axis fixed to the QD
(the choice of which drops out of the calculation).
The tunneling through junction $r$ is assumed to conserve spin and occurs with a spin-independent amplitude  $t_r$,
thereby setting the tunnel rates $\Gamma_r = 2\pi \sum_\tau \nu_{r\tau} t_r^2$.
The spin dependence of tunnel amplitude in $H_T$,
$t_{r \sigma \tau} =
\bra{\sigma} e^{-i \frac{1}{2} \vecg{\chi}_r \cdot \vecg{\sigma}} \ket{\tau}
 t_r $,
arises because we use a different quantization axis in each part of the system.
Here $\chi_r=|\vecg{\chi}_r|$ is the angle of rotation about the vector $\vecg{\chi}_r$,
which maps the QD $z$-axis onto the polarization vector $\vec{n}_r$.
The electrodes are held at temperature $T$ and
 the transport is controlled (i) by biasing the electro-chemical potentials $\mu_r= \pm V_b/2$ of the electrodes with $V_b$,
(ii) by controlling the level position $\epsilon = - V_g $ through the gate voltage $V_g$,
and (iii) by adjusting the relative polarization angle $\theta$ 
 ($\vec{n}_L \cdot \vec{n}_R = |\vec{n}_L| |\vec{n}_R| \text{cos}~\theta$).
\par
\emph{Transport quantities and continuity equations}.
The theory we now develop can address the important question
how the impurity spin in the above model can be detected by charge transport
 and controlled by the applied voltages.
To understand how charge, spin and SQM can accumulate on the high-spin QD,
we first derive the associated current operators and continuity equations.
The change of the number operator $\hat{N}$ for electrons localized on the QD
is induced by the injected electron particle currents,
$
\hat{I}_{N}^{r} = -i [H_\text{T},\hat{N}^{r}]
$
where
$
\hat{N}^r  =  \sum_{k\tau} c_{r k \tau}^{\dagger}c_{r k \tau}
$ is the electron number on FM $r$:
\begin{align}
  \dot{\hat{N}}  = \sum_r \hat{I}_{N}^{r}
  \label{eq:Ncont}
  .
\end{align}
This follows from the conservation of the total charge of the system,
$
\hat{N}^\text{tot} = \hat{N}+ \sum_r \hat{N}^r
$
,
in a tunneling process,
$
[H_T, \hat{N}^\text{tot}] = 0
$,
and the conservation of charge on the QD without tunneling,
$
[H, \hat{N}] = 0
$.
A similar consideration for the spin operators shows that the change in the QD spin $\veco{S}$ is generated entirely by the spin-conserving tunneling if the QD is spin isotropic, i.e.
$
[H, \veco{S}] = 0
$:
\begin{align}
  \dot{\veco{S}} = \sum_r \veco{I}_\vec{S}^{r}
  \label{eq:Scont}
\end{align}
with spin-current operators
$
\hat{I}_{\vec{S}}^{r} = -i [H_\text{T},\veco{S}^{r}]
$
and spin polarization
$
\veco{S}^r  = \fracs{1}{2} \sum_{k\tau \tau'} c_{r k \tau}^{\dagger}\vecg{\sigma}_{\tau \tau'} c_{r k \tau'}
$
for FM $r$.
\begin{figure}[t]
  \includegraphics[width=0.7\linewidth]{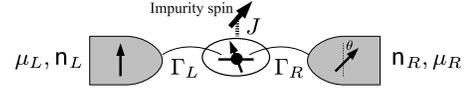}
  \caption{Quantum dot high-spin valve: noncollinear ferromagnets coupled to an orbital level probing a spin impurity.}
  \label{fig:setup}
\end{figure}
The central aspect of the high-spin QD  of interest here is that the average of the local quadrupole tensor \eq{eq:Q} can become nonzero:
the spin triplet state is then anisotropic in addition to spin polarized.
This anisotropy is induced by the \emph{SQM currents} entering the QD from the spin-anisotropic environment:
\begin{align}
  \dot{\tenso{Q}} =  \sum_r \tenso{I}_{\tens{Q}}^{r} 
  \label{eq:Qcont}
  .
\end{align}
This SQM continuity equation is be obtained by using the conservation of the total SQM of the system in the tunneling, which, however, involves various canceling contributions.
Instead, the SQM current operator associated with electrode $r$,
$
 \tenso{I}_{\tens{Q}}^{r} =
 \fracs{1}{2}
 (
   \veco{I}_\vec{S}^r \veco{S} + \veco{S} \veco{I}_\vec{S}^r 
 )
 - ( \fracs{1}{3}  \veco{I}_\vec{S}^r \cdot \veco{S} ) \tenso{1}
 + \text{h.c}.
$,
can be obtained more easily from definition \eq{eq:Q}, using the product rule for the time-derivative and the spin continuity equation \eq{eq:Scont}.
Since $\braket{ \veco{I}_\vec{S}^r \veco{S} } \neq \braket{ \veco{I}_\vec{S}^r} \braket{\veco{S} }$,
the expectation value of the SQM-current in \eq{eq:Qcont} has no simple relation to the spin current.
To calculate the accumulation of charge, spin and SQM on the QD
we need to microscopically derive a theory which describes the QD nonequilibrium state determined by the competition between the charge, spin and SQM currents discussed above.
\par
\emph{Kinetic equations.}
The time-evolution of the (reduced) density operator $p$ of the QD, describing its nonequilibrium state, is determined by the QD Hamiltonian $H$ and a kernel in the kinetic (generalized master) equation.
We diagrammatically calculate the time-evolution kernel to the first order in $\Gamma_r$ in the framework of the real-time transport theory.
To describe the QD high-spin valve in the single-electron tunneling (SET) regime,
while consistently accounting for quantum coherence~\cite{Braun04set,*Donarini09},
all degrees of freedom  need to be considered, including the eight nondiagonal elements of the QD density matrix in the eigen basis of $H$.
Without making additional approximations we rewrite the resulting kinetic equations as an exactly equivalent set of coupled equations for quantum-statistical \emph{averages} of a complete set of physical operators in which the density operator can be expanded.
For our model these are
the charge occupancies $p^{d}$, $p^{t}$ ($p^{d}+p^{t}=1$) of the doublet and triplet state respectively
($N = \Tr \hat{N}  p  = p^{t}$)
,
the corresponding spin accumulations $\vec{S}^{d}$, $\vec{S}^{t}$
($\vec{S} = \Tr \veco{S} p  = \vec{S}^{d}+\vec{S}^{t}$)
and
the triplet SQM $\tens{Q}^{t}$
($\tens{Q} = \Tr \tenso{Q} p = \tens{Q}^{t}$).
Note that in a QD high-spin valve the spin accumulation vectors $\vec{S}^{d}$, $\vec{S}^{t}$ in the two accessible charge states  differ in orientation and magnitude and need to be calculated separately.
Deferring the further technical details to elsewhere, we discuss the physical meaning of the resulting equations in the stationary limit:
\begin{align}
  0=  \dot{p}^{d}   = &  
  -3          {\gamma}^{+}            {p}^{d}  +          2        {\gamma}^{-}           {p}^{t}
  -2     \vecg{\gamma}^{+} \cdot \vec{S}^{d}   +          2   \vecg{\gamma}^{-} \cdot \vec{S}^{t}
  \label{eq:pDdot}
  \\
  0=  \dot{p}^{t}   = &
  \phantom{-}
  3          {\gamma}^{+}            {p}^{d}   -          2        {\gamma}^{-}           {p}^{t}
  +
  2     \vecg{\gamma}^{+} \cdot \vec{S}^{d}   -          2   \vecg{\gamma}^{-} \cdot \vec{S}^{t}
  \label{eq:pTdot}
  \\
  0=\dot{\vec{S}}^{d}  = & 
  - \fracs{1}{2}  \vecg{\gamma}^{+}          {p}^{d} +  \fracs{1}{3} \vecg{\gamma}^{-}           {p}^{t}
  -
  3           {\gamma}^{+}      \vec{S}^{d}   +                    {\gamma}^{-}       \vec{S}^{t}
   + 
    \vec{S}^{d} \times \vecg{\beta}
  \nonumber \\ &
    +
    2  \,    \tens{Q}^{t} \cdot \vecg{\gamma}^{-}
    \label{eq:SDdot}
  \\
  0=\dot{\vec{S}}^{t}  = & 
  \phantom{-} \fracs{4}{2}  \vecg{\gamma}^{+}          {p}^{d}   - \fracs{4}{3} \vecg{\gamma}^{-}           {p}^{t}
  +
  4           {\gamma}^{+}      \vec{S}^{d}   -   2               {\gamma}^{-}       \vec{S}^{t}
  + 
     \vec{S}^{t} \times \vecg{\beta}
  \nonumber \\ &
  -  2 \,    \tens{Q}^{t} \cdot \vecg{\gamma}^{-}
  \label{eq:STdot}
  \\
  0=\dot{\tens{Q}}^{t}  = & 
  \left[
  \fracs{4}{2} \left( \vec{S}^{d} \vecg{\gamma}^{+} +  \vecg{\gamma}^{+}\vec{S}^{d} \right)
  - \fracs{4}{3} (\vec{S}^{d} \cdot \vecg{\gamma}^{+} ) \tens{1}
  \right]
  \nonumber \\ &
  -
  \left[
  \fracs{1}{2} \left( \vec{S}^{t} \vecg{\gamma}^{-} + \vecg{\gamma}^{-} \vec{S}^{t}  \right)
  - \fracs{1}{3} (\vec{S}^{t} \cdot \vecg{\gamma}^{-} ) \tens{1}
  \right]
  \nonumber \\ &
  -  2 \gamma^{-} \tens{Q}^{t} 
  + \tens{Q}^{t} \times \vecg{\beta}
  -\vecg{\beta} \times \tens{Q}^{t}
  \label{eq:QTdot}
\end{align}
Here
$     {\gamma}^{\pm}=\sum_r      {\gamma}_r^{\pm}$
and
$
\gamma_r^{\pm} = \fracs{1}{2} \Gamma_r f^{\pm}_r(\epsilon)
$
is the rate for single charge tunneling in/out of the QD through junction $r$,
denoting the Fermi distribution for electrons / holes by
$f_r^{\pm}(\epsilon) = ( e^{\pm( \epsilon-\mu_r )/ T } + 1 )^{-1}$.
Similarly,
bold-faced vectors $\vecg{\gamma}^{\pm}=\sum_r \vecg{\gamma}_r^{\pm}$
and
$
\vecg{\gamma}_r^{\pm} = \gamma_r^{\pm} \vec{n}_r 
$
denote the corresponding rates of tunneling of spin, polarized along $\vec{n}_r$, through junction $r$.
Finally, $\vecg{\beta}=\sum_r \vecg{\beta}_r$ is an effective magnetic field,
with contributions from each electrode $r$:
\begin{align}
  \vecg{\beta}_r = 
  \fracs{1}{2} \Gamma_r \mathrm{Re}
  \int \limits_{-D}^{D} \frac{d \omega}{\pi} \frac{f_r^{+}(\omega)}{\omega-\epsilon+i0}
  \vec{n}_r
\label{eq:exchange}
\end{align}
where the cut-off needs to be set to $D \sim U$ since we exclude the $N=2$ charge state of the model.
This field represents the spin splitting induced on the QD by \emph{coherent} virtual electron tunneling processes into the spin-polarized electrode $r$ and relies on the nonzero value of the Coulomb charging energy $U$~\cite{Koenig03}.
The magnitude $|\vecg{\beta_r}|$ of these exchange fields  is electrically tunable~\cite{Hauptmann08}, with a peak at $\epsilon=\mu_r$ and logarithmic tails.
The kinetic equations incorporate the conservation of probability, $p^{d} + p^{t}=1$, and the tracelessness of the SQM tensor, $\text{tr}\,\tens{Q}^t= \sum_i Q_{ii}^t=0$.
The stationary charge current through junction $r$ can be calculated in a similar way:
\begin{align}
  I_N^r   = 
  3          {\gamma}^{+}_r            {p}^{d}   -          2        {\gamma}^{-}_r           {p}^{t}
  +
  2     \vecg{\gamma}^{+}_r \cdot \vec{S}^{d}   -          2   \vecg{\gamma}^{-}_r \cdot \vec{S}^{t}
  \label{eq:I}
  .
\end{align}
The continuity equation \eq{eq:Ncont} is satisfied:
$ \dot{N} = \dot{p}^{t} = - \dot{p}^{d} = \sum_r I_N^r = 0 $ in the stationary limit.

Due to the spin polarization of the electrodes, the charge occupancies \eq{eq:pDdot}-\eq{eq:pTdot} and the current \eq{eq:I} couple to both charge-specific spins, but are  not directly influenced by the SQM.
Eqs.~\eq{eq:SDdot}-\eq{eq:STdot} show that these spins couple back to the charge occupancies,
and suffer isotropic spin relaxation
($\propto -\gamma^{+}\vec{S}^d$ and $\propto -\gamma^{-}\vec{S}^t$, respectively).
In addition there is  a transfer of spin polarization from one charge state to the other
($\propto \gamma^{-}\vec{S}^t$ and $\propto \gamma^{+}\vec{S}^d$, respectively).
The next-to-last term represents a torque  on the charge-specific spin due to the exchange field $\vecg{\beta}$.

A central result of this Letter is that the spin accumulated in each charge state also couples to the SQM accumulated in the triplet state through the last term $\pm 2 \tens{Q}^t \cdot \vecg{\gamma}^{-}$ in \eq{eq:SDdot}, \eq{eq:STdot}.
The opposite signs of these terms indicate that a nonzero SQM 
tends to make the spin polarizations in the two charge states noncollinear.
The accumulation of SQM is described by the kinetic equation \eq{eq:QTdot}:
the net injection (first two terms, using dyadic notation $(\vec{a}\vec{b})_{ij}=a_i b_j$) competes with the isotropic relaxation (third term).
The last two terms,
$(\tens{Q}^t \times \vecg{\beta})_{i j} = \epsilon_{jkl}Q^t_{ik}\beta_{l}$
and
$(\vecg{\beta} \times \tens{Q}^t)_{i j} = \epsilon_{ikl}\beta_{k}Q^t_{lj}$
,
incorporate a torque exerted on the SQM by the exchange field $\vecg{\beta}$.
This finite SQM results in a back-action on the spin
when substituted into the right hand side of \eq{eq:SDdot}-\eq{eq:STdot}.
Importantly, these back-action terms
are comparable to the other terms in \eq{eq:SDdot}-\eq{eq:STdot}.
By solving \Eq{eq:QTdot} for $\tens{Q}^t$ in terms of $\vec{S}^{d}$ and $\vec{S}^{t}$
they can be written as
\begin{align}
  2 \, \vec{Q}^{t} \cdot \vecg{\gamma}^{-} =
  \sum_{\lambda=d,t} \left(
    \tens{R}^{\lambda} \cdot \vec{S}^{\lambda}
    + \vec{S}^{\lambda} \times \vecg{\beta}^{\lambda}
  \right)
  \label{eq:back}
  .
\end{align}
The first two terms in \eq{eq:back} make the spin relaxation and the spin transfer \emph{anisotropic} through symmetric tensors $\tens{R}^\lambda$.
The remaining terms have the form of a spin torque involving new, \emph{charge-state specific exchange fields} $\vecg{\beta}^{\lambda}$ which can be shown to be \emph{noncollinear} with the standard exchange field $\vecg{\beta}$.
These have the two-fold effect
 of  modifying the existing spin torque term in Eqs.~\eq{eq:SDdot}-\eq{eq:STdot},
$
\vecg{\beta} \rightarrow \vecg{\beta} \pm \vecg{\beta}^{d,t}
$,
and  adding a torque which involves the spin from the \emph{other charge state}.
The latter thus represents a {transfer of spin torque} between the two charge states.
In contrast to the standard exchange field $\vecg{\beta}$, which is of a purely coherent origin (cf.~\eq{eq:exchange}), both $\vecg{\beta}^{\lambda}$ and $\tens{R}^\lambda$ arise from a complex interplay of dissipative and coherent processes:
their (lengthy) expressions contain both the transition rates $\vecg{\gamma}^{\pm}$ and the exchange field $\vecg{\beta}$.
We emphasize that the above spin relaxations and spin torques in \eq{eq:back} cannot be understood as arising from spin (or charge) currents:
for a correct description of the spin dynamics the transport of SQM must be accounted for.
By its effect on the spin, the SQM also acts back on the charge occupancies and the measurable charge current, cf. \Eq{eq:I}.
From an exhaustive study of all parameter regimes of the model we find that
 in general the SQM significantly affects the charge transport whenever spin accumulation occurs.
To emphasize its impact we note that
if one neglects the SQM in Eqs.~\eq{eq:pDdot}-\eq{eq:QTdot}
when the drain electrode is most strongly spin polarized,
one obtains an unphysical, large particle current running opposite to the voltage bias direction
for nearly all relative polarization angles $\theta$.
We now return to the questions raised for the studied model.
\begin{figure}[t]
    \includegraphics[width=0.95\linewidth]{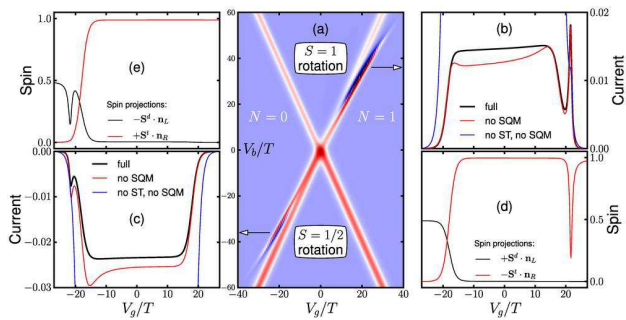}
  \caption{
    (Color online).
    (a) Stability diagram $dI/dV_b$ vs. $V_g$, $V_b$
    (red = positive, dark blue = negative).
    Parameters:
    $\Gamma_\text{L} = 2 \Gamma_\text{R} = 0.2T$
    and $D= U=500T$ in units of the temperature $T$,
    and $\theta = 0.99 \pi$.
    The high polarization $|\vec{n}_L|=|\vec{n}_R|=0.99$ allows
    all effects to be clearly illustrated,
    lower values only result in a rescaling of the current plateaus.
    (b),(c): $I$ vs. $V_g$ for $V_b=\pm 36 T$ normalized to the maximal current
    $I_0 =((3\Gamma_{L,R})^{-1}+ (2\Gamma_{R,L})^{-1})^{-1}$  achievable for $V_b \gtrless 0$
    and parallel polarizations ($\theta=0$).
    Black: full result,
    red: neglecting the SQM,
    blue: additionally neglecting the spin-transfer (ST) terms in \Eq{eq:SDdot}-\eq{eq:STdot}.
    (d),(e): Spin projections
    $\pm\vec{S}^{d}\cdot \vec{n}_L$ and
    $\mp\vec{S}^{t}\cdot \vec{n}_R$ affecting the current $\eq{eq:I}$ vs. $V_g$
    corresponding to (b) and (c), respectively.    
  }
  \label{fig:result}
\end{figure}
\par
\emph{Impurity spin detection and control.}
When similar FMs ($|\vec{n}_L|=|\vec{n}_R|$) with \emph{nearly anti-parallel} polarizations ($\theta \approx \pi$) are asymmetrically tunnel coupled ($\Gamma_L > \Gamma_R$) to the QD,
the presence of the side-coupled impurity spin can be detected by  nonlinear transport measurements.
\Fig{fig:result}(a-c) show that a sharp, anomalous current peak occurs in the thermally broadened regime where usually the SET current through the QD switches on.
The peak height depends nonmonotonically on $V_b$ and its position depends nonlinearly on $V_g$.
The occurrence of such a current peak for opposite polarities of the bias and gate voltage (relative to the degeneracy point)
indicates the presence of ferromagnetic coupling of the orbital to an impurity spin,
resulting in nonzero total spin in both accessible charge states.
Indeed, a calculation which ignores the impurity spin completely, only shows the current peak for forward bias.
\Fig{fig:result}(d)-(e) show that in both cases the current peak directly measures a significant precession of the spin accumulation on the QD.
This strong precession in a narrow range of voltages is possible due the asymmetric tunneling coupling.
At the bias and gate voltage where the effect is maximal
the magnitude of the field $\vecg{\beta}_R(\epsilon)$ originating form the weakly coupled FM
is resonantly enhanced ($|\epsilon-\mu_R| \approx T$)
and matches the magnitude of $\vecg{\beta}_L(\epsilon)$ from the strongly coupled FM
which is off resonance ($|\epsilon-\mu_L| \gg T$).
For $\theta \sim \pi$ the exchange field $\vecg{\beta}$ is a sum of the two nearly anti-parallel, equal-length vectors \eq{eq:exchange}
and
is therefore perpendicular to the spin, which accumulates in one of the QD charge states opposite to either $\vec{n}_R$ or $\vec{n}_L$ for $\mu_L \gtrless \mu_R$, see \Fig{fig:result}(d),(e).
 Although this exchange field is small, it does, however, cause a large spin precession over an angle $\sim$ $\pi$ due to the onset of Coulomb blockade
which
suppresses the spin relaxation.
When the weakly coupled ferromagnet is tuned off resonance with $V_g$ or $V_b$, $|\epsilon-\mu_R| \gg T$, this precession is switched off.
The correct description of this electrically controlled spin resonance relies on both the spin transfer terms (cf. blue curve) as well as the SQM induced back-action on the spin (cf. red curve).
\par
\emph{Conclusion.} The illustrated interplay of nonequilibrium transport of charge, spin polarization and spin anisotropy in high-spin nanostructures may open up new possibilities for electrical detection and manipulation of spin~\cite{Sothmann10,Kiesslich09}.
We believe that ``spin-quadrupoletronics'' -- the transport of spin anisotropy --   presents interesting challenges and important prospects for further theoretical and experimental work.
For instance, the design of experimental setups which support pure SQM currents (not accompanied by spin currents) is of interest for creating spin reversal barriers in spin-isotropic systems.
For QDs with spins $S > 1$ the theory can be extended to transport of spin-multipole moments of rank $2S$ required to describe higher degrees of spin anisotropy.
We acknowledge stimulating discussions with
B. Sothmann,
J. K\"onig,
and S. Andergassen.
and the financial support from
NanoSci-ERA and DFG (FOR 912).
\bibliographystyle{apsrev}

\end{document}